\documentstyle[12pt,fleqn]{article}
\begin{document}

\begin{center}
{\bf Microwave-induced Resonant Reflection and Localization of Ballistic
Electrons in Quantum Microchannels}\footnote{
Submitted to Surface Science, Proceedings of EP2DS XI,
Nottingham, August 1995}
\end{center}
\begin{center}
L.~Y.~Gorelik$^\ast$, M.~Jonson, and R.~I.~Shekhter\\
Chalmers University of Technology and G\"oteborg University, S-412
96 G\"oteborg,  Sweden, and
$^\ast$B. Verkin Institute for Low Temperature Physics and Engineering,
Kharkov, Ukraine.
\end{center}
\begin{abstract}
We show that electron transport in a ballistic microchannel supporting both
propagating and reflected modes can be completely blocked by applying a
microwave electromagnetic field. The effect is due to resonant reflection
caused by multiple coherent electron-photon scattering involving at least
two spatially localized scattering centers in the channel.
With many such scattering centers present the conductance
is shown to have an irregular dependence on bias voltage, gate voltage and
frequency with irregularily spaced dips corresponding to resonant reflection.
When averaged over bias, gate voltage or frequency the conductance
will decay exponentially with channel length in full analogy with the
localization of 1D electrons caused by impurity scattering.
%
\end{abstract}
The dynamics of a mesoscopic system subject to a time dependent field
depends cruically on the relation between the phase breaking time
$\tau_\phi$ and the time $t_0$ needed for electrons to pass through the
system. In the absence of phase breaking processes, $\tau_\phi \gg t_0$,
one could expect phase coherent dynamic phenomena to occur in
mesoscopic systems in close analogy with the well known static phenomena.
They should be highly tunable by for instance a magnetic field or by electric
fields due to applied bias- or gate voltages. The photoconductance of a
ballistic microchannel created in a gated AlGaAs heterostructure is an
example. In a channel with a cross section that varies so slowly that the
adiabatic approximation applies, coherent electron-photon scattering
corresponding to transitions between different transverse modes has been
shown to be localized in space and to lead to transitions between
propagating and reflected modes \cite{Nazarov}. This type of indirect
backscattering has many features in common with impurity scattering and
has been suggested \cite{ourprl} to give rise to quantum interference
effects in the electron transport properties. In particular the
photoconductance caused by single photon scattering (absorption) was shown
to be an oscillating function of an applied gate voltage in a quasi-one
dimensional channel containing a microwidening \cite{ourprl}. At large
enough electromagnetic fields, multiple coherent electron-photon scattering
results in a coherent resonant coupling of two different transverse modes
in the channel. Because the electron-photon interaction is localized in
space, this resonant coupling can be expressed in terms of a Landau-Zener
breakdown \cite{LandauZener}. In this paper we pursue this line of argument
and show that resonant electron-photon scattering strongly modifies the
electron transport and that it may lead to resonant (total) reflection of
electrons and hence block the electron transport through a quasi-1D
channel completely.

The channel geometry we have in mind is shown in Fig.~1a. We consider the
simplest case where only a single transverse mode is ballistically
propagating through the channel (schematically illustrated by the bold
lines in Fig.~1a). The microwave-induced resonant coupling of
this propagating mode with localized modes in the smooth widenings of
the channel (thin lines in Fig.~1a) is the phenomen of interest here. Still
referring to Fig.~1a, the longitudinal coordinates $x_1(n)$ and $x_2(n)$
give the positions where the resonant condition $\Delta U \equiv U_2(x) -
U_1(x) =\hbar\omega$ is satisfied and Landau-Zener scattering takes place.
Here $U_n(x)$ is the effective potential barrier for the quasi-one
dimensional motion of electrons in the $n$:th transverse mode
\cite{Nazarov}. The total transmission through the channel is determined
by the sum of the probability amplitudes for all electron paths. In the
straight parts of the channel shown in Fig.~1a the electron states are
superpositions of left- and right moving plane waves of amplitudes
$A_1(n)$ and $A_2(n)$ respectively. Each widening is characterized by a
scattering matrix ${\bf S}_n$, which couples the amplitudes of incoming
and outgoing waves,
\begin{equation}
\left(
\begin{array}{c} A_1(n+1) \\ A_2(n) \end{array}
\right)
= {\bf S}_n
\left(
\begin{array}{c} A_1(n) \\ A_2(n+1) \end{array}
\right), \quad
{\bf S}_n
\equiv {\rm e}^{i\xi_n}
\left(
\begin{array}{cc}
i{\cal T}_n & {\cal R}_n^\ast \\
{\cal R}_n & i{\cal T}_n
\end{array}
\right).
\end{equation}

We will first calculate the scattering matrix ${\bf S}_n$ and then later
use the result to calculate the total transmission probability of the
channel, which determines the conductance through the Landauer formula.

\begin{figure}
\vspace{3,5cm}
\caption{(a) Schematic view of a quasi-1D channel with one propagating
mode (thick arrows) and several microwidenings with trapped modes (thin
arrows).
(b) Diagram describing scattering between propagating and trapped modes in a
microwidening (see text)
}  \label{f.1}
\end{figure}

The internal structure of the scattering matrix ${\bf S}_n$ is
schematically shown in Fig.~1b (where the index $n$ is suppressed for
convenience). Scattering is localized in space and may occur at $x_1$ or
$x_2$. At each site scattering of right moving electrons (upper part of
Fig.~1b) and left moving electrons (lower part) must be considered
separately. The closed curve in the middle of Fig.~1b represents the path
of an electron state trapped in the widening. The scattering matrices
${\bf S}_{ic}$ and ${\bf S}_{id}$ that appear in Fig.~1b describe
Landau-Zener transitions at point $x_i$ for the case when the
transverse energy levels are converging ($c$) and diverging ($d$) in the
direction of longitudinal motion. They are related as
${\bf S}_{id}=\sigma_y {\bf S}_{ic}\sigma_y$, where $\sigma_y$ is a
Pauli spin matrix, and can be expressed in terms of two phases $\theta_i$,
$\vartheta_i$ --- related to the amplitudes of the two resonantly coupled
modes --- and the probability $r_i^2$ for an intermode transition as
\begin{equation}
\label{Sic}
{\bf S}_{ic}
\equiv {\rm e}^{i\theta_i}
\left(
\begin{array}{cc}
t_i & ir_i {\rm e}^{i\vartheta_i} \\
ir_i {\rm e}^{-i\vartheta_i} & t_i
\end{array}
\right), \quad t_i^2 + r_i^2 =1 , \quad
r_i = \left(1-{\rm e}^{-\pi a_i}\right)^{1/2}.
\end{equation}
The quantity $r_i$ is determined
by the single parameter
$a_i = e^2 {\cal E}^2/p\omega |\Delta U^\prime(x_i)|$ (as are $\theta_i$,
$\vartheta_i$). ${\cal E}$ and $\omega$ is the amplitude and frequency of
the electromagnetic field polarized in the transverse direction,
while $p$ is the
longitudinal momentum of the electron at the scattering sites.

There is a useful formal analogy between our scattering problem and the
familiar
case of electron scattering by (tunneling through) a double barrier structure.
If we interpret $A_1(n)$ and $A_1(n+1)$ as the amplitudes of an incident and a
reflected wave, then ${\bf S}_{1c}$ and ${\bf S}_{2d}$ correspond to the
first and
${\bf S}_{1d}$ and ${\bf S}_{2c}$ to the second barrier in the analogous
double barrier
tunneling problem.

Note that the two barriers are identical, since each has one part
corresponding to scattering at site $x_1$ and another part due to scattering at
$x_2$. This immediately tells us that there is a possibibility for
reflectionless transmission from $A_1(n)$ through the double barrier structure
to $A_2(n)$ for certain values of the electron energy, which we can readily
calculate. In the original problem, however, $A_2(n)$ is the amplitude of the
{\em reflected} rather than the transmitted wave (cf. Fig.~2). Hence resonant
coupling of two transverse modes by an electromagnetic field can lead to
resonant reflection of electrons at a microwidening in an otherwise straight
channel.

Straightforward calculations lead to the following result for the transmission
probability through the microwidening:
\begin{equation}
\label{T2E}
{\cal T}^2(E) = \frac{
4 T^2\sin^2\left(\phi_L(E)-\eta\left(\Delta\varphi(E)\right)\right)
}{
\left(1-T^2\right)^2 +
4T^2\sin^2\left(\phi_L(E)-\eta\left(\Delta\varphi(E)\right)\right)
}
\end{equation}
Here the phase $\phi_L$ is the total phase gained by the electron as it
propagates around the closed trajectory (middle part of Fig.~1b), while
$\Delta\varphi$ is the difference in phase gained along the two
different trajectories going from scatterer  ${\bf S}_{1c}$  to ${\bf
S}_{2d}$. These two phases depend on two independent geometric parameters, say
the length and width of the microwidening. In an experiment that implies they
depend on applied gate voltages and on radiation frequency.
While the phase $\eta(\varphi)$ in (\ref{T2E}) is always less than $\pi/2$, the
phase $\phi_L$ is a linear function of energy in the quasiclasical limit so
that a discrete set of electron energies $\{E_m\}$ exists for which
$\sin(\phi_L(E_m)-\eta(\Delta\varphi(E_m)))=0$. At these energies the
transmission probability ${\cal T}^2$ goes to zero and resonant reflection
takes place. The width of the resonance is determined by the parameter $T^2$,
which can be expressed in terms of the probability $r^2$ for
inter-mode Landau-Zener transitions. In the simple case of a symmetric widening
($r_1=r_2=r$), we find
\begin{equation}
T^2 = 1 - 4 t^2 r^2 \sin^2\left(\Delta \varphi/2\right)
\end{equation}
It is interesting to note that $T^2$ is close to unity and the effect of the
electromagnetic field is small for the two limiting cases of weak ($t\ll
1$) and strong ($r\ll 1$) field as long as we are outside the resonant regions
(which are narrow in both cases). Corrections are due to
the fact that $1-T^2 \ne 0$ and coincide with those calculated in the single
photon approximation \cite{ourprl}. This approximation breaks down at the
resonant energies, however, where the transmission is suppressed to zero
irrespective of the field strength. For a single microwidening in an otherwise
straight channel the conductance
$
G = (2e^2/h) {\cal T}^2\left(E=E_F\right)
$
should therefore show a series of resonance ``dips'' as the gate voltage is
varied. This resonant structure becomes richer and less regular if several
microwidenings are present. The transmission amplitude can in this case be
obtained from a product of transfer matrices describing the individual
widenings. If the numer of widenings is large --- and if there are no
correlations between the parameters decsribing their geometries --- the theory
of disordered one-dimensional systems can be used to find the result
\cite{Thouless}
\begin{equation}
G = \frac{2e^2}{h} \exp\left(\sum_{n=1}^N\ln|{\cal T}_n|\right) .
\end{equation}
Here ${\cal T}_n$ is the transmission amplitude for the $n$:th widening as
given by (\ref{Sic}). The criterion ${\cal T}_n=0$ for at least one $n$
determines the gate voltages for which electron transport is completely blocked
by the electromagnetic field. Taking an ensemble average in the standard theory
of disordered systems corresponds here to averaging over gate voltage. Hence we
get for our problem the same exponential decay of conductance found for the
electron localization problem in disordered (equilibrium) conductors, i.e.
\begin{equation}
G = \frac{2e^2}{h}  {\rm e}^{-N/R_L}, \quad
R_L^{-1} = \left\langle \ln {\rm Max}(r_n,t_n)\right\rangle, \quad
\left\langle f_n\right\rangle \equiv \lim_{N\to\infty} \frac{1}{N}
\sum_{n=1}^N f_n
\end{equation}
Note that the localization radius $R_L$ depends on the probability for
Landau-Zener transitions,

The prospects for observing the effects discussed above in an experiment
depend on the probability for intermode scattering caused by other mechanisms,
such as electron-phonon or impurity scattering. Two different criteria apply
for the possibilities to observe resonant reflection and the exponential
suppression of the averaged current, respectively. In the first case the rate
$\nu$ of non-radiative inter-mode transitions should be smaller than the width
of the discrete electron spectrum of trapped (non-propagating) modes in a
widening:
$
\nu \ll r^2 t^2\omega (d/L_W)
$.
Here $L_W$ is the characteristic length of the widening and we have used
the relation $v_F\sim \omega d$ between the Fermi velocity and channel width.

If we assume that $\nu$ is due to phonon emission we have $\nu\sim\gamma
\omega^3/\omega_D^2$ --- $\omega_D$ is the Debye frequency and $\gamma$
the electron-phonon coupling constant --- and conclude
that the inequality
\begin{equation}
\gamma\left(\omega/\omega_D\right)^2 \ll r^2 t^2
\left(d/L_W\right)
\end{equation}
must be fullfilled for resonant reflection to be observable. This should be
possible to achieve experimentally.
For the exponential decay of the averaged conductance to be observable, it is
necessary to maintain phase coherence along the entire channel (not only within
one microwidening). Therefore the stronger inequality
$\gamma(\omega/\omega_D)^2\ll
(d/L)$
must be fullfilled. Here $L$ is the total length of the channel and we have
assumed $r \sim t \sim 1$.

This work was supported by the Swedish Institute, the
Swedish Engineering Science Research Council (TFR), and by Swedish
Natural Science Research Council (NFR).

\end{document}